\documentclass
[aps,prl,twocolumn,floatfix,english,showpacs,10pt,superscriptaddress]{revtex4-2}%
\usepackage{graphicx}
\usepackage{booktabs}
\usepackage{amsmath}
\usepackage{physics}
\usepackage{amssymb}
\usepackage{colordvi}
\usepackage{verbatim}
\usepackage{xcolor}
\usepackage{mathrsfs}
\usepackage{epsfig}
\usepackage{lipsum}
\usepackage{amsfonts}
\usepackage{makecell}
\usepackage{esint}
\usepackage{bm}
\usepackage{ulem}
\usepackage[most]{tcolorbox}

\usepackage[unicode=true, breaklinks=false, pdfborder={0 0 1}, backref=false,
colorlinks=true, linkcolor=blue, urlcolor=blue, citecolor=blue]{hyperref}%
\providecommand{\U}[1]{\protect\rule{.1in}{.1in}}
\providecommand{\U}[1]{\protect\rule{.1in}{.1in}}
\setcitestyle{numbers,square}

\begin{document}

\title{Non-Relativistic Spin-Orbit Interaction in Triplet Superconductors: Edelstein Effect and Spin Pumping by Electric Fields}

\author{Ping Li}
\affiliation{School of Physics, Huazhong University of Science and Technology, Wuhan 430074, China}

\author{G. A. Bobkov}
\affiliation{Moscow Institute of Physics and Technology, Dolgoprudny, 141700 Moscow region, Russia}

\author{I. V. Bobkova}
\email{ivbobkova@mail.ru}
\affiliation{Moscow Institute of Physics and Technology, Dolgoprudny, 141700 Moscow region, Russia}

\author{Tao Yu}
\email{taoyuphy@hust.edu.cn}
\affiliation{School of Physics, Huazhong University of Science and Technology, Wuhan 430074, China}

\date{\today}

\begin{abstract}
Non-relativistic momentum-dependent spin splitting, as observed in collinear altermagnets and non-collinear $p$-wave magnets, provides exciting avenues for controlling spin dynamics. Here, we reveal a distinct form of non-relativistic ``spin–orbit coupling" in triplet superconductors by demonstrating that the triplet order parameter induces a wave-vector-dependent spin texture of Bogoliubov quasiparticles, thereby entangling their orbital and spin motions. Even in the absence of relativistic spin–orbit coupling, this intertwining of spin and orbital motion allows an electric field to generate spin polarization in a $p$-wave superconductor---that is, an Edelstein effect. Building on this mechanism, we propose an efficient scheme for the nonlinear generation of a DC spin current via electric near fields, driven by AC spin polarization and electron velocity. This general principle offers a powerful route for generating and manipulating spin currents in unconventional superconductors.

\end{abstract}

\maketitle

\textit{Introduction}.---Spin-orbit coupling (SOC) plays a crucial role in numerous condensed matter systems, such as topological phases~\cite{topological_phase_1,topological_phase_2,topological_phase_3,topological_phase_4,topological_phase_5}, magnetism~\cite{DMI_1,DMI_2,DMI_3,DMI_4}, spintronic devices~\cite{spintronics_1,soc2,spintronics_3}, and quantum computing platforms~\cite{quantum_computing}. As a relativistic effect in electrons, the spin degree of freedom entangles with orbital motion, leading to the conversion of charge and spin currents and to electric/optical control of magnetization~\cite{spintronics_1,soc2,spintronics_3}. Momentum-dependent spin splitting is a key consequence of relativistic SOC, but recent studies demonstrate spin splitting that does not rely on relativistic effects~\cite{AM1,AM2,AM3,OFe,LBai,CSong}. Altermagnets exhibit even-parity collinear magnetic order and display momentum-dependent spin splitting in the absence of a net magnetization~\cite{AM1,AM2,AM3,AM4,AM5}. $p$-wave magnets, characterized by odd-parity non-collinear magnetic order~\cite{AM3,LBai,p_wave_magnet1,SHa,p_wave_magnet3}, also exhibit strong non-relativistic spin-momentum coupling, providing a non-relativistic analog to the Rashba or Dresselhaus SOC~\cite{Rashba1960,Dresselhaus1955,review_5,p_wave_magnet1}. Remarkable functionalities of non-relativistic SOC include spin-anisotropic transport, the potential for charge-to-spin conversion~\cite{charge_spin4, p_wave_magnet3,charge_spin2,charge_spin3}, the non-relativistic Edelstein effect~\cite{p_wave_magnet4,p_wave_magnet5,spin_Edelstein3}, and the spin-galvanic effect~\cite{p_wave_magnet6,spin_galvanic2}.

\begin{figure}[htp!]	
\centering
\centering
\includegraphics[width=1\linewidth, clip, trim=0.5cm 4cm 2cm 1cm]{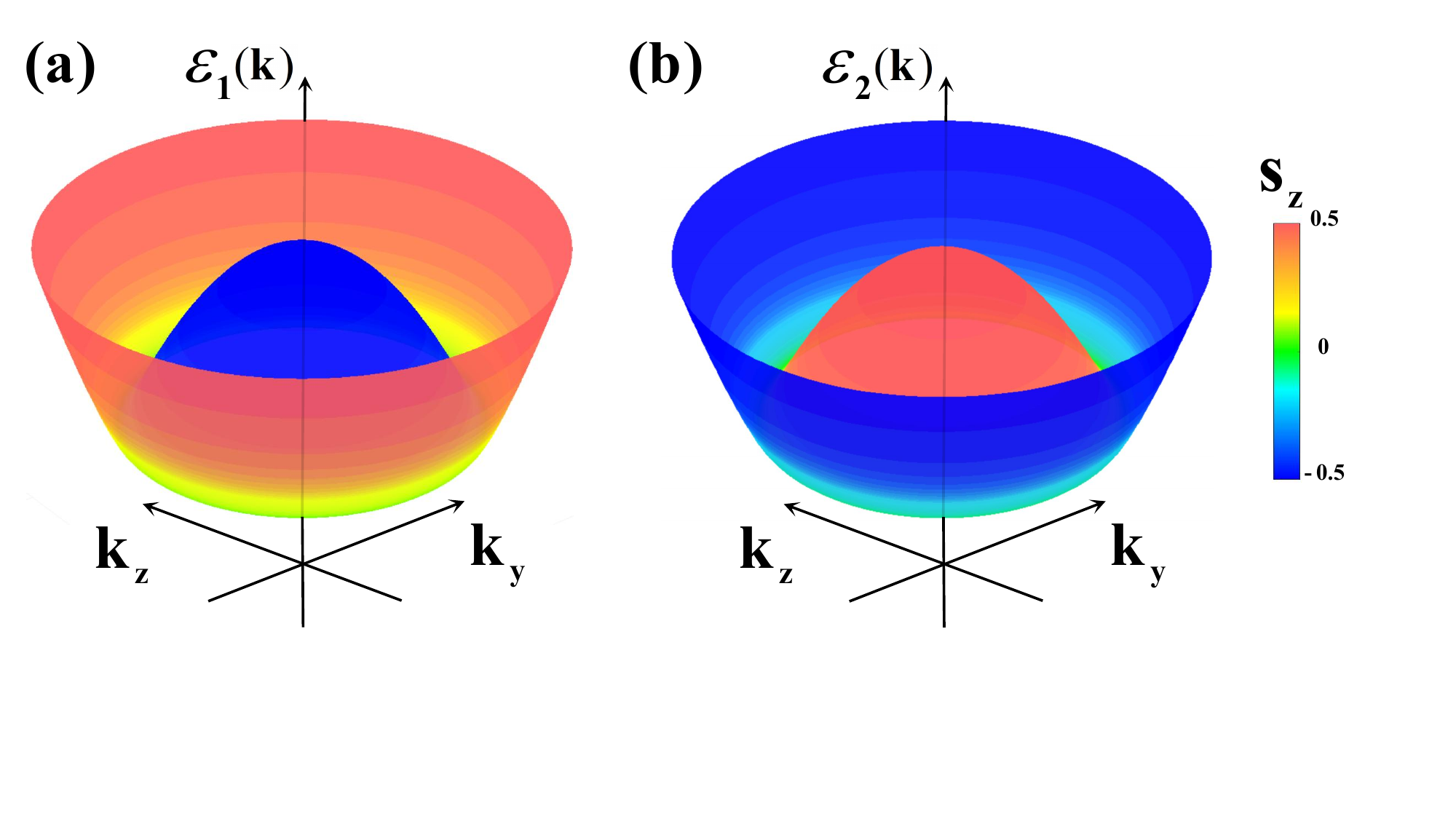}
\caption{Illustration of spin $S_z\in[-\hbar/2,\hbar/2]$ of Bogoliubov quasiparticles on the two degenerate bands $\{\varepsilon_1({\bf k}),\varepsilon_2({\bf k})\}$ of a triplet superconductor in the momentum space. The color maps represent the spin of quasiparticles $S_z$.  $S_z$ depends strongly on the wave vector near the Fermi wave vector $k_F$.}
\label{scattering_process}
\end{figure}

In this Letter, we discover a different type of non-relativistic SOC that emerges in triplet superconductors, in which, although the quasiparticle dispersion is spin-degenerate, the spin of the Bogoliubov quasiparticle is strongly wave-vector-dependent near the Fermi wave vector, as illustrated in Fig.~\ref{scattering_process}, which acts as an effective coupling between the spin and orbital motion. Unlike the standard SOC, inversion symmetry in the non-superconducting state remains intact and is broken solely by the triplet order parameter $\hat \Delta_t(\bm{k}) = -\hat \Delta_t(-\bm{k})$. Triplet superconducting order parameter is expected to occur in different superconductors~\cite{JYang,JKBao,KMTaddei,Huakunzuo,ShengRan,TMetz,DAoki,JD,HTou,MKriener,MYokoyama,interface,C_Liu,Z_Chen,Hanwei}. Based on this effective SOC, an AC electric field enables spin polarization in triplet superconductors, i.e., realizing the spin Edelstein effect~\cite{Edelstein_effect1,Edelstein_effect2,Edelstein_effect3,Edelstein1995,Edelstein2005,Bobkova2016} free of relativistic SOC. This effect distinguishes itself from the generation of orbital magnetization by irradiating a superconductor with a circularly polarized electromagnetic field~\cite{orbital1, orbital2,Plastovets2022,Mironov2025,orbital3}. Furthermore, based on this effect, we propose the spin pumping~\cite{spin_pumping} solely driven by the near electric field~\cite{metallic_nano1,metallic_nano2,Y_Au,chirality_theorem,nano_optics,plasmonics,Betzig,LeWang,Vincent,HaominWang}. Indeed, the electric field plays a crucial role in nonlinear responses: the velocity driven by the electric field combines with the spin polarization generated by the Edelstein effect, thereby efficiently producing a DC spin current flowing outward from the driven region. 


\textit{Non-relativistic SOC in triplet superconductors}.---We illustrate the general principle with a minimal model of a triplet $p$-wave superconducting film, taking the surface normal along the $\hat{\bf x}$-direction, as realized, for example, at the LaAlO$_3$/KTaO$_3$ interface~\cite{interface,C_Liu,Z_Chen,Hanwei}. In Nambu space, the Hamiltonian matrix
\begin{align}
{\cal H}_0({\pmb \rho})=
\left(\begin{array}{cccc}
\frac{\hbar^2{\hat {\bf k}}^2}{2m}-\mu & 0 & \Delta_t e^{i \hat \theta_{\bf k}} & 0 \\
0 & \frac{\hbar^2{\hat {\bf k}}^2}{2m}-\mu & 0 & -\Delta_t e^{-i \hat \theta_{\bf k}} \\
\Delta_t e^{-i\hat \theta_{\bf k}}& 0 & -(\frac{\hbar^2{\hat {\bf k}}^2}{2m}-\mu) & 0 \\
0 & -\Delta_t e^{i\hat \theta_{\bf k}} & 0 & -(\frac{\hbar^2{\hat {\bf k}}^2}{2m}-\mu)\end{array}
\right),
\label{Hamiltonian}
\end{align}
in which ${\pmb \rho}=y\hat{\bf y}+z\hat{\bf z}$, $\hat {\bf k} = -i \nabla_{\bm \rho}$, $m$ is the electron mass, $\mu$ is the electron chemical potential, $\Delta_t$ is the magnitude of the triplet order parameter, and $e^{i \hat \theta_{\bf k}} = (\hat k_y + i \hat k_z)/|{\bf k}|$ is its phase. In terms of the field operator $\hat{\Psi}({\pmb \rho})=(\hat{\psi}_\uparrow(\pmb \rho),\hat{ \psi}_\downarrow(\pmb \rho),\hat{\psi}^\dagger_\uparrow(\pmb \rho),\hat{ \psi}^\dagger_\downarrow(\pmb \rho))^T$, the Hamiltonian $\hat{H}_0=(1/2)\int d{\pmb \rho}\hat{\Psi}^\dagger({\pmb \rho}){\cal H}_0({\pmb \rho})\hat{\Psi}({\pmb \rho})$. The spin density operator $\hat{\bf s}({\pmb \rho})=(\hbar/4)\hat{\Psi}^{\dagger}({\pmb \rho}){\pmb{\cal S}}\hat{\Psi}({\pmb \rho})$ evolves according to the continuity equation
$\partial_t\hat{\bf s}({\pmb \rho})+\pmb\nabla\cdot\hat{\bf J}_s({\pmb \rho})=\hat{\bf T}_s({\pmb \rho})$, where   
${\pmb{\cal S}}=\left(\begin{array}{cc}{ \pmb\sigma}  &  \\
& -{\pmb\sigma}^\ast
\end{array}\right)$ 
represents the Pauli matrices in Nambu space, the spin-current density operator
\begin{align}
    \hat{\bf J}_{ s}({\pmb \rho})=\frac{i\hbar^2}{8m}\hat{\Psi}^\dagger({\pmb \rho})(\overleftarrow{\pmb{\nabla}}-\overrightarrow{\pmb{\nabla}})\tau_3{\pmb{\cal S}}\hat{\Psi}({\pmb \rho}),
    \label{spin_current_operator}
\end{align}
and the spin-torque density operator
\begin{align}
    \hat{\mathbf{T}}_s({\pmb \rho})&=\frac{\Delta_t}{4i}\hat{\Psi}^\dagger({\pmb \rho})\Big[\left(\begin{array}{cccc}
         0&0&{\overleftarrow{e^{i\theta_{\bf k}}}}&0  \\
         0&0&0&-\overleftarrow{e^{-i\theta_{\bf k}}}\\
         \overleftarrow{e^{-i\theta_{\bf k}}}&0&0&0\\
         0&-\overleftarrow{e^{i\theta_{\bf k}}}&0&0
    \end{array}\right){\pmb{\cal S}}\nonumber\\
    &+{\pmb{\cal S}}\left(\begin{array}{cccc}
         0&0&{\overrightarrow{e^{i\theta_{\bf k}}}}&0\\
         0&0&0&-\overrightarrow{e^{-i\theta_{\bf k}}}\\
         \overrightarrow{e^{-i\theta_{\bf k}}}&0&0&0\\
         0&-\overrightarrow{e^{i\theta_{\bf k}}}&0&0
    \end{array}\right)\Big]\hat{\Psi}({\pmb \rho}).
    \label{spin_torque_density_operator}
\end{align} 
Here, $\tau_3={\rm diag}(1,1,-1,-1)$ is the metric in Nambu space. 
The spin density $\hat{\bf s}({\pmb \rho})$ of electrons is thereby not conserved in the presence of the triplet order parameter $\Delta_t$, even in the absence of the SOC.

Based on the Hamiltonian~\eqref{Hamiltonian}, the quasiparticle dispersion, given by $\varepsilon_{1}(k)={\varepsilon}_{2}(k)=\sqrt{\xi_k^2+\Delta_t^2}$ and ${\varepsilon}_{3}(k)={\varepsilon}_{4}(k)=-\sqrt{\xi_k^2+\Delta_t^2}$, with $\xi_k={\hbar^2k^2}/({2m})-\mu$, are spin degenerate. However, the triplet order parameter induces spin mixing in each state individually (see Supplemental Material (SM)~\cite{supplement}). Unlike conventional SOC, this mixing occurs in the {\it particle-hole channel}: the quasiparticle wavefunction comprises an electron component $u_{\uparrow(\downarrow)}$ and a hole component $v_{\uparrow(\downarrow)}$, the latter representing the absence of an electron with spin $\uparrow(\downarrow)$, i.e., a hole excitation with spin $\downarrow(\uparrow)$. The degree of mixing is set by the triplet order parameter, $(v_\uparrow/u_\uparrow)_{\alpha=1,3} = \pm \Delta_t e^{-i\theta_k}/(\varepsilon_\alpha(k) \pm \xi_k)$ and $(v_\downarrow/u_\downarrow)_{\alpha=2,4} = \mp \Delta_t e^{i\theta_k}/(\varepsilon_\alpha(k) \pm \xi_k)$. This particle-hole spin mixing renders the spin carried by Bogoliubov quasiparticles momentum dependent, $s_{\alpha=1,2}(k) = \pm \xi_k/\varepsilon_{\alpha}(k)$, as shown by the color map in Fig.~\ref{scattering_process}---in sharp contrast to the singlet case, where every quasiparticle has spin projection $\pm 1$ for any quantization axis. The triplet order parameter thus acts as a {\it non-relativistic SOC for electron and hole quasiparticles}.

\textit{Scattering theory}.---The non-relativistic SOC renders nontrivial spin response in superconductors to the electric field. A local monochromatic electromagnetic field of frequency $\omega$, when applied to the superconductor film of area $A$, generally reads 
\[
\left\{{\bf E}({\boldsymbol \rho},t),{\bf H}({\boldsymbol{\rho},t})\right\}=\frac{1}{A}\sum_{\bf k}\sum_{\zeta=\pm} \left\{{\bf E}^\zeta({\bf k}),{\bf H}^\zeta({\bf k})\right\}   e^{-i\zeta\omega t}e^{i{\bf k}\cdot{\boldsymbol\rho}},
\]
which is readily produced by sources including the near electromagnetic fields from excited nanostructures (metallic, ferroelectric, or magnetic)~\cite{metallic_nano1,metallic_nano2,chirality_theorem}, focused sub-THz optical radiation~\cite{nano_optics,plasmonics,Betzig,LeWang,Vincent,HaominWang}, or a scanning near-field optical microscope (SNOM)~\cite{near_field_optic1,near_field_optic2,near_field_optic3,near_field_optic4}.

The electron orbitals couple to the electric field ${\bf E}({\pmb \rho},t)=-\partial {\bf A}({\pmb \rho},t)/\partial t$ or the vector potential ${\bf A}({\pmb \rho},t)$ through $\hat{H}_E(t)=\int d{\pmb \rho}\hat{\Psi}^{\dagger}(\pmb \rho){\cal H}_E(\pmb \rho,t)\hat{\Psi}(\pmb \rho)$, where, disregarding the diamagnetic ${\bf A}^2$-term, ${\cal H}_E({\pmb \rho},t)=-({e}/{4m}) 
    [{\bf A}({\pmb \rho},t)\cdot {\bf p}+{\bf p}\cdot{\bf A}({\pmb \rho},t)]{\cal I}_{4\times 4}$. The electron spin density $\hat{\bf s}(\pmb \rho)$ interacts with the AC magnetic field ${\bf H}({\pmb \rho},t)$ via the Zeeman interaction $\hat{H}_H(t)=\mu_0\gamma\int d{\pmb{\rho}}\hat{\bf s}(\pmb \rho)\cdot {\bf H}(\pmb \rho,t)$, 
with $\mu_0$ the vacuum permeability and $\gamma$ the gyromagnetic ratio of electrons. 
The matrix $\Phi({\bf k}) =(\varphi_1({\bf k}),\varphi_2({\bf k}),\varphi_3({\bf k}),\varphi_4({\bf k}))$ and annihilation operators of quasiparticles $\hat{{\pmb \gamma}}_{\bf k}=(\hat{\gamma}_{{\bf k}1},\hat{\gamma}_{{\bf k}2} ,
\hat{\gamma}^\dagger_{-{\bf k}1},\hat{\gamma}^\dagger_{-{\bf k}2})^T$ expand the field operator as
$\hat{\Psi}({\pmb \rho})=({1}/{\sqrt{A}})\sum_{\bf k}\Phi({\bf k}) e^{i{\bf k}\cdot{\pmb \rho}}\hat{\pmb\gamma}_{\bf k}$.
Including couplings to electric and magnetic fields, the interaction Hamiltonian
\begin{align}
    \hat{H}_{\rm int}(t)=\sum_{{\bf k}{\bf k}^\prime}\sum_{\zeta=\pm}\sum^4_{\alpha,\beta=1}{G}^\zeta_{\alpha\beta}({\bf k},{\bf k}^\prime)\hat{\gamma}^\dagger_{{\bf k}\alpha}\hat{\gamma}_{{\bf k}^\prime\beta}e^{-i\zeta\omega t},
    \label{interection_Hamiltonian}
\end{align}
where the elements in the $4\times 4$ matrix 
\begin{align}
{G}^\zeta({\bf k},{\bf k}^\prime)&=\frac{i\zeta e\hbar}{4m\omega A}({\bf k}+{\bf k}^\prime)\cdot{\bf E}^\zeta({\bf k}-{\bf k}^\prime)\Phi^\dagger({\bf k})\Phi({\bf k}^\prime)\nonumber\\
&+\frac{\mu_0\gamma\hbar}{4A}{\bf H}^\zeta({\bf k}-{\bf k}^\prime)\cdot \Phi^\dagger({\bf k})\pmb{\cal S}\Phi({\bf k}^\prime)
\label{G_matrix_EH}
\end{align}
are the coupling constants between the Bogoliubov quasiparticles and electric and magnetic fields.

The external electromagnetic field in Eq.~\eqref{interection_Hamiltonian} acts as a local time-dependent scattering potential for the incident quasiparticles. 
The operator $\hat{\pmb \Gamma}_{{\bf k}\alpha}=\sum_{{\bf q}'}\sum_{\alpha'=1}^4T_{\alpha\alpha'}({\bf k},{\bf q}',t)\hat{\gamma}_{{\bf q}'\alpha'}$
combines the incident quasiparticles operator $\hat{\gamma}_{{\bf q}'\alpha'}$
 with the scattering amplitudes encoded in the $T$-matrix, whose elements
\begin{align}
    &T_{\alpha'\alpha}({\bf q}',{\bf k},t)=\delta_{{\bf q}'{\bf k}}\delta_{\alpha'\alpha}\nonumber\\
    &+\sum_{\zeta=\pm}U_{\alpha'\alpha}^{\zeta}({\bf q}',{\bf k})\exp\left[-i\left(\zeta\omega-\omega_{\alpha'}(q')+\omega_\alpha(k)\right)t\right]\nonumber\\
    &+\sum_{\zeta_1\zeta_2}Q_{\alpha'\alpha}^{\zeta_1\zeta_2}({\bf q}',{\bf k}) \exp\left[-i(\omega_{\alpha}(k)+(\zeta_1+\zeta_2)\omega-\omega_{\alpha'}(q'))t\right].\nonumber
\end{align}
Here, $\omega_{\alpha}(k)={\varepsilon}_{\alpha}(k)/\hbar$ are the quasiparticle frequencies.
The scattering amplitudes for a single photon process---where the electron absorbs ($\zeta=-$) or emits ($\zeta=+$) one photon---are given by the complex quantities
\begin{align}
U_{\alpha'\alpha}^{\zeta}({\bf q}',{\bf k})=\frac{{G}^\zeta_{\alpha'\alpha}({\bf q}',{\bf k})}{\zeta\hbar\omega-{\varepsilon}_{\alpha'}(q')+{\varepsilon}_\alpha(k)+i\delta},
\end{align}
with $\delta\rightarrow0^+$ introduced from the adiabatic turning-on of the interaction.  The scattering amplitudes for two-photon processes are 
\begin{align}
    Q_{\alpha'\alpha}^{\zeta_1\zeta_2}({\bf q}',{\bf k})&
    =\sum_{{\bf q}'',\alpha''}\frac{{G}^{\zeta_1}_{\alpha'\alpha''}({\bf q}',{\bf q}'')}{{\varepsilon}_{\alpha''}(q'')-{\varepsilon}_\alpha(k)-\zeta_2\hbar\omega-i\delta}\nonumber\\
    &\times \frac{{G}^{\zeta_2}_{\alpha''\alpha}({\bf q}'',{\bf k})}{{\varepsilon}_{\alpha'}(q')-{\varepsilon}_\alpha(k)-(\zeta_1+\zeta_2)\hbar\omega-i\delta},
    \label{scattering_amplitudes_2}
\end{align}
where the two-photon sequences are represented by $\{\zeta_1,\zeta_2\}=\{+,-\}$ (first emission, then absorption) or $\{-,+\}$ (first absorption, then emission).

Under the scattering of the electromagnetic field, the quasiparticle field operator is modified to
\begin{align}
    \hat{\Psi}({\pmb \rho},t)=\frac{1}{\sqrt A}\sum_{\bf k}{\Phi}({\bf k})e^{i{\bf k}\cdot{\pmb \rho}}\hat{\pmb \Gamma}_{\bf k}(t).
    \label{field_operator_scattering}
\end{align} 
Substituting this into the spin-current density operator \eqref{spin_current_operator} gives the DC spin-current density for $\hbar \omega<2\Delta_t$:
\begin{align}
&{\bf J}^{\rm DC}_{s}({\pmb \rho})    
=\frac{\hbar^2}{4mA}\sum_{{\bf k}{\bf k}^\prime{\bf q}}\sum_{\zeta=\pm}\sum^2_{\alpha\beta\eta=1}\left({\bf M}_{\beta\alpha}({\bf k},{{\bf  k}^\prime})\otimes{\bf k}^\prime\right)\nonumber\\
&\times{G}^{-\zeta}_{\alpha\eta}({\bf k}^\prime,{\bf q}){G}^{\zeta}_{\eta\beta}({\bf q},{\bf k})e^{i({\bf k}^\prime-{\bf k})\cdot{\pmb \rho}}\nonumber\\ 
&\times\frac{f({\varepsilon}_{\beta}(k))-f({\varepsilon}_{\eta}(q))}{\left({\varepsilon}_{\eta}(q)-{\varepsilon}_{\alpha}( k^\prime)-\zeta\hbar\omega+i\delta\right)\left({\varepsilon}_{\beta}( k)-{\varepsilon}_{\eta}(q)+\zeta\hbar\omega+i\delta\right)}\nonumber\\
&+{\rm H.c.},
\label{DC_spin_current}
\end{align}
where the matrix elements are defined as
${\bf M}_{\alpha\beta}({\bf k},{\bf k}^\prime)=\varphi_\alpha^\dagger({\bf k})\tau_3{\pmb {\cal S}}\varphi_\beta({\bf k}^\prime)$. This contribution arises from processes involving photon absorption followed by emission, or the reverse sequence. Furthermore, substituting the field operator \eqref{field_operator_scattering} into the spin-torque density operator \eqref{spin_torque_density_operator} yields the DC spin-torque density induced by the AC field (see SM~\cite{supplement} for details).

\begin{figure}[htp!]
\centering
\includegraphics[width=1\linewidth, clip, trim=4cm 1cm 4cm 0cm]{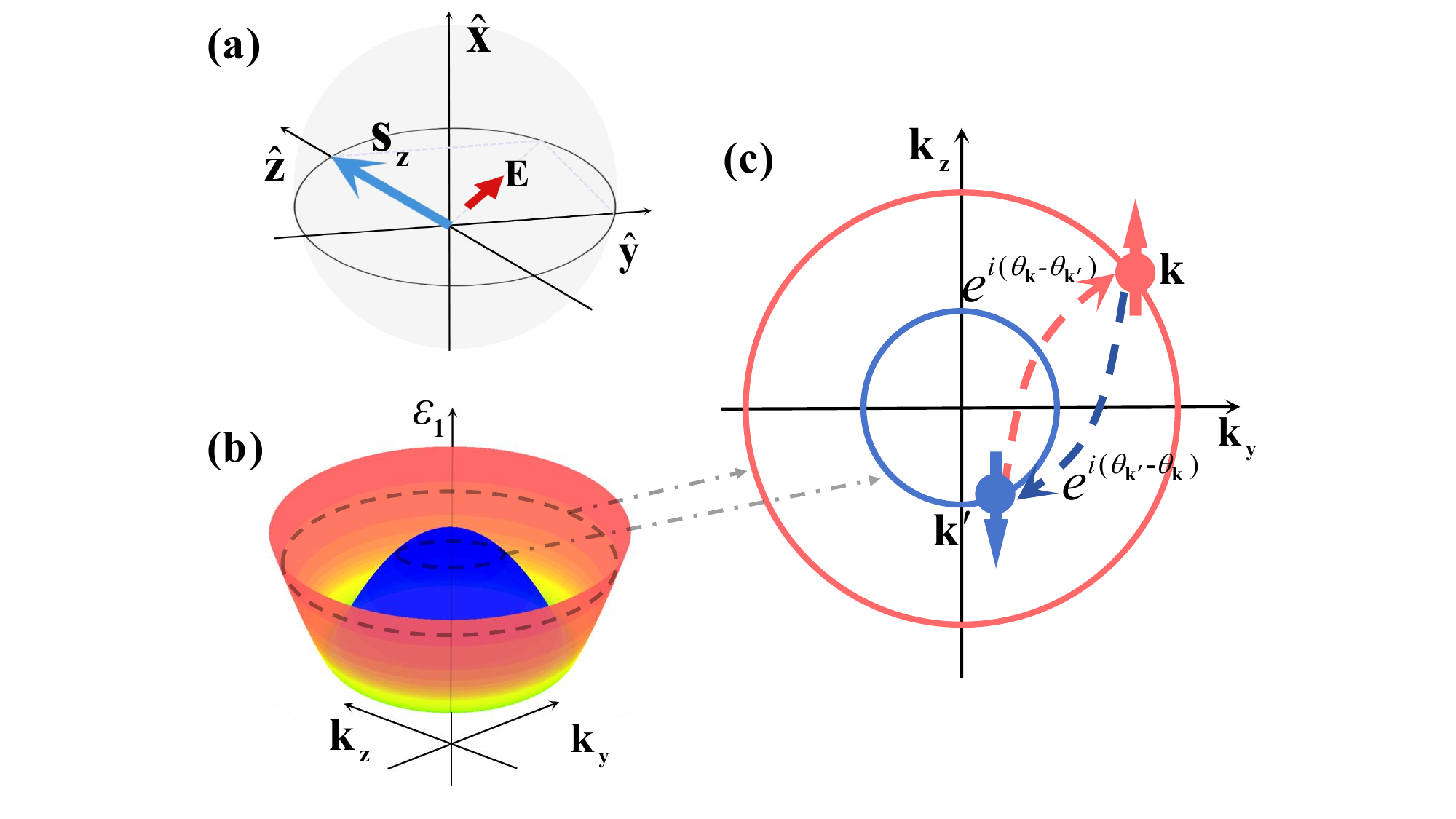}
\caption{Spin Edelstein effect in a $p$-wave superconductor. (a): an in-plane electric field ${\bf E}$  (red arrow) induces the spin density polarized along the $\hat{\bf z}$-direction (blue arrow). (b) $\&$ (c): This spin polarization arises from an intra-band quasiparticle scattering process driven by the electric field, as illustrated for band $1$ and addressed in the main text.}
\label{Intra_Band_Scattering}
\end{figure}

\textit{Spin Edelstein effect}.---Due to the non‑relativistic SOC, the electric field plays a significant role in the spin response of a triplet $p$-wave superconductor, even though it does not couple directly to the electron spin. Analyzing  the linear response of the spin density to the  electromagnetic field $\{{\bf E}({\pmb\rho},t),{\bf H}(\pmb\rho,t)\}$  offers a direct way to examine the induced spin polarization:
\begin{align}
    &\delta{\bf s}({\pmb\rho},t)=\frac{\hbar}{4A}\sum_{{\bf k}{\bf k}^\prime}\sum_{\zeta=\pm}\sum_{\alpha,\beta=1,2}{\pmb{\cal S}}_{\beta\alpha}({\bf k},{\bf k}^\prime){ G}_{\alpha\beta}^\zeta({\bf k}^\prime,{\bf k})\nonumber\\
    &\times\frac{e^{i[({\bf k}^\prime-{\bf k})\cdot{\pmb\rho}-\zeta\omega t]}}{\zeta\hbar\omega-{\varepsilon}_{\alpha }(k^\prime)+{\varepsilon}_{\beta}( k)+i\delta}\left[2f({\varepsilon}_{\beta}(k))-1\right]+{\rm H.c.},\nonumber
    \end{align}
where the matrix element ${\pmb{\cal S}}_{\beta\alpha}({\bf k},{\bf k}^\prime)=\varphi^\dagger_\beta({\bf k}){\pmb{\cal S}}\varphi_\alpha({\bf k}^\prime)$ represents the spin operator acting between quasiparticle states $\beta$ and $\alpha$.
 For an in-plane electric field, the coupling constant $G^{\zeta,E}_{\alpha\beta}({\bf k}^\prime,{\bf k})=\delta_{\alpha\beta}G^{\zeta,E}_{\alpha\alpha}({\bf k}^\prime,{\bf k})$ is diagonal in the eigenbasis of \eqref{Hamiltonian}.  
 Consequently, the spin polarization induced by the in-plane electric field arises solely from intra-band processes, and---since $\pmb{\cal S}_{\beta\alpha}({\bf k},{\bf k}^\prime)\delta_{\alpha\beta}=(0,0,{\cal S}^z_{\alpha\alpha}({\bf k},{\bf k}^\prime))$---it generates spin polarization only along the $\hat{\bf z}$-direction, as illustrated in Fig.~\ref{Intra_Band_Scattering}(a). The resulting $z$-component of the spin density is given by $\delta{\bf s}_z({\pmb\rho},t)\propto\sum_{{\bf k}{\bf k}^\prime}\sin(\theta_{\bf k}-\theta_{{\bf k}^\prime})({\bf k}+{\bf k}^\prime)\cdot{\bf E}^\zeta({\bf k}^\prime-{\bf k})+{\rm H.c.}.$

The spin Edelstein effect in the $p$-wave superconductor arises from two key ingredients: (i) the wave-vector-dependent quasiparticle spin, which acts as an effective coupling between spin and orbital motion. (ii) the momentum-dependent phase in the spin-triplet pairing potential, which enables antisymmetric quasiparticle scattering processes under the electric field.   Figure~\ref{Intra_Band_Scattering}(b,c) illustrates the intra-band quasiparticle scattering process driven by the electric field for the band $\alpha=1$. The blue (red) arrow represents an electron-to-hole (hole-to-electron) conversion that decreases (increases) $s_z$. This process is accompanied by the creation (annihilation) of a Cooper pair with spin up. The two opposite spin-generation processes do not cancel each other because in this band, the hole-like component of the quasiparticle wavefunction carries a momentum-dependent phase  $v_{\uparrow}({\bf k})\propto e^{-i\theta_{\bf k}}$, while the electron-like component $u_\uparrow({\bf k})$ is phase-independent. Consequently, as addressed in Fig.~\ref{Intra_Band_Scattering}(c), the optical transition from an electron-like quasiparticle at wavevector ${\bf k}$ on the outer energy contour (red  circle) to a hole-like quasiparticle at ${\bf k}'$ on the inner contour (blue circle) acquires a phase factor
$e^{i(\theta_{{\bf k}^\prime}-\theta_{\bf k})}$. 
The reverse process, from ${\bf k}'$ back to ${\bf k}$, yields the complex-conjugate phase  $e^{i(\theta_{{\bf k}}-\theta_{{\bf k}^\prime})}$ and an opposite change in spin. The interference between these two scattering paths therefore produces a finite spin polarization $s_z$ proportional to $\sin(\theta_{\bf k}-\theta_{{\bf k}^\prime})$. A similar intra-band scattering process takes place in band $\alpha=2$, where the hole-like component $v_\downarrow(\bf k)$ carries the phase $e^{i\theta_{\bf k}}$. The process in the two bands likewise contributes a net spin polarization of the same sign.

\textit{Spin pumping by electric field}.---Substituting the coupling constant  Eq.~\eqref{G_matrix_EH}, we obtain the DC spin-current density and DC spin-torque density pumped by the electromagnetic field. The effect of the electromagnetic field can be decomposed into three contributions: a pure magnetic-field term $\sim {\bf H}^2$, which is negligibly small; a combination of electric and magnetic fields $\sim \{{\bf E},{\bf H}\}$; and a pure electric field $\sim {\bf E}^2$.

The electric field drives the electron velocity $\delta {\bf v}(t)$ primarily along the field direction. This velocity, combined with the excited spin polarization $\delta {\bf s}_z(t)$, gives rise to a DC spin current $\sim \delta{\bf s}_z(t)\otimes \delta{\bf v}(t)$ polarized along the $\hat{\bf z}$-direction. Consequently, in the triplet $p$-wave superconductor described by Eq.~\eqref{Hamiltonian}, all in-plane components of the electric field induce DC spin currents and spin torques polarized solely along the $\hat{\bf z}$-direction,  expressed as $\sum_{i,j=y,z}g_{ij}E_i^*E_j$ in terms of the response coefficients coefficients $g_{ij}$.  The magnetic field also polarizes the particle spin, while the electric field drives the electron velocity. Therefore, the combination of electric and magnetic fields is expected to generate DC spin currents and spin torques via terms of the form  $\sum_{ij}f_{ij}{ E}_i^*{ H_j}$ with response coefficients $f_{ij}$.  The pumped spin-current ${\bf J}_s^{x,y}$ and spin-torque ${\bf T}_s^{x,y}$ densities receive contributions from both $\sum_{i=y,z}f_{ix}{ E}_i^*{ H}_x$ and $\sum_{i=y,z}f_{iy}{ E}_i^*{ H}_y$; meanwhile, the pumped spin-current ${\bf J}_s^{z}$ and spin-torque ${\bf T}_s^{z}$ densities take the form $\sum_{i=y,z}f_{iz}{ E}_i^*{ H}_z$.  All possible configurations of the DC spin-current and spin-torque densities generated via second-order response in the $p$-wave superconductor—classified by the direction of spin polarization, the direction of flow, and the type of field combination—are summarized in Table~\ref{tab:selection-rules} (see SM~\cite{supplement} for details).
 
\begin{table}[htp!]
\centering
\caption{Contributions from electromagnetic coupling to the DC spin-current ${\bf J}_s$ and spin-torque ${\bf T}_s$ densities in a prototypical $p$-wave superconductor. The superscripts on ${\bf J}_s$ and ${\bf T}_s$ indicate the spin-polarization direction.  Here, $g_{ij}$ and $f_{ij}$ denote the corresponding response coefficients.}
\label{tab:selection-rules}
\begin{tabular}{@{}ll@{}}
\toprule
\textbf{spin current/torque} &~~~~~~~ \textbf{governed by}\\
\midrule
${\bf J}_{y,z}^{z}$,~~~ ${\bf T}_s^{z}$ &$\sum_{i=y,z}g_{ij}E_i^*E_j$\\
${\bf J}_{y,z}^{x,y}$,~~~ ${\bf T}_s^{x,y}$ &$\sum_{i=y,z}f_{ix} E_i^* H_x$, $\sum_{i=y,z}f_{iy}E_i^* H_y$\\
${\bf J}_{y,z}^{z}$,~~~ ${\bf T}_s^{z}$ &$\sum_{i=y,z}f_{iz}E^*_i H_z$\\
\bottomrule
\end{tabular}
\end{table}

\begin{figure}[htp!]
    \centering
     \centering
    \includegraphics[width=0.95\linewidth, clip, trim=2cm 2cm 1cm 3cm]{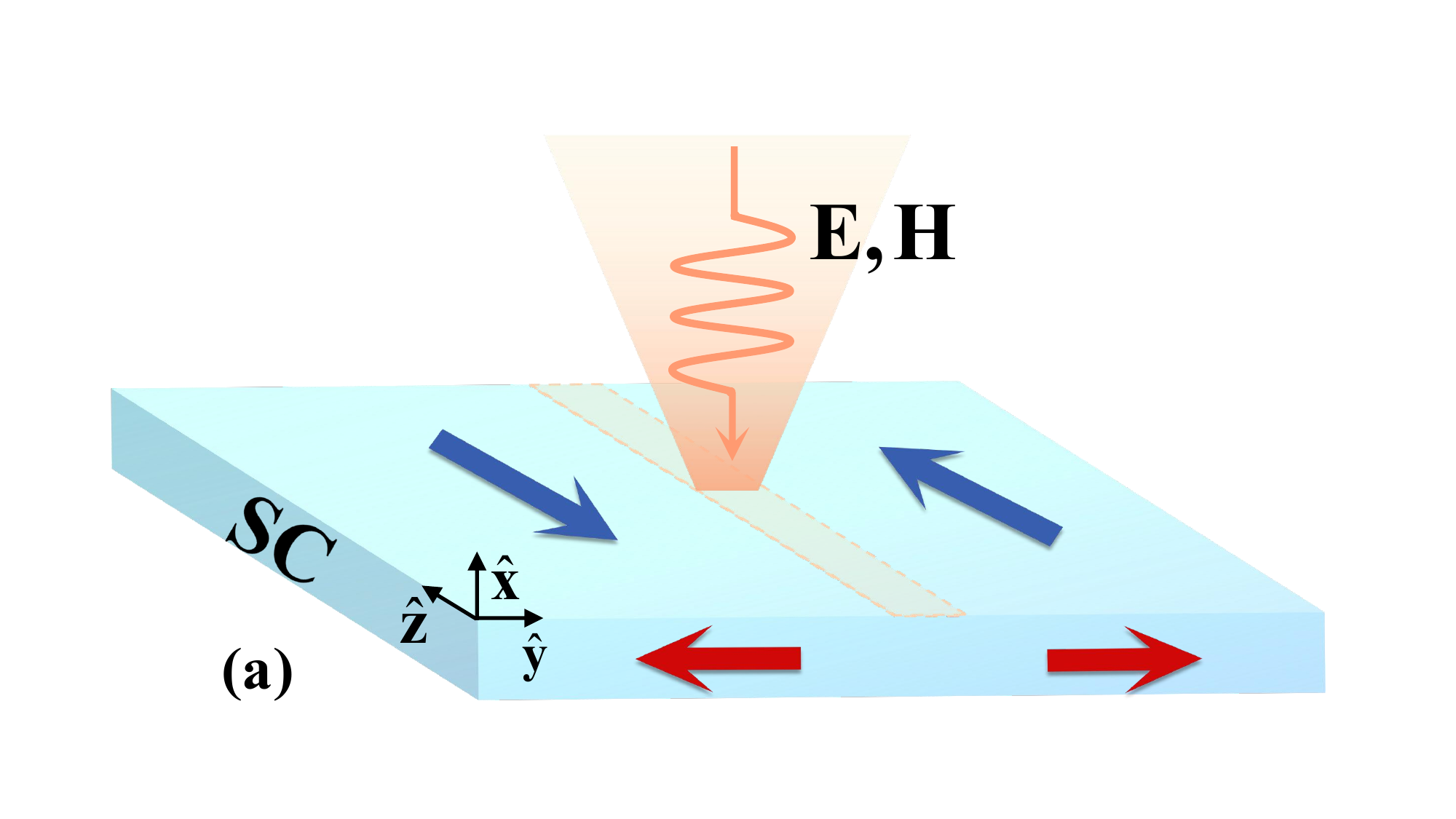}
    \begin{minipage}[b]{0.492\columnwidth}
        \centering
        \includegraphics[width=\linewidth, clip, trim=1cm 0cm 0.1cm -1cm]{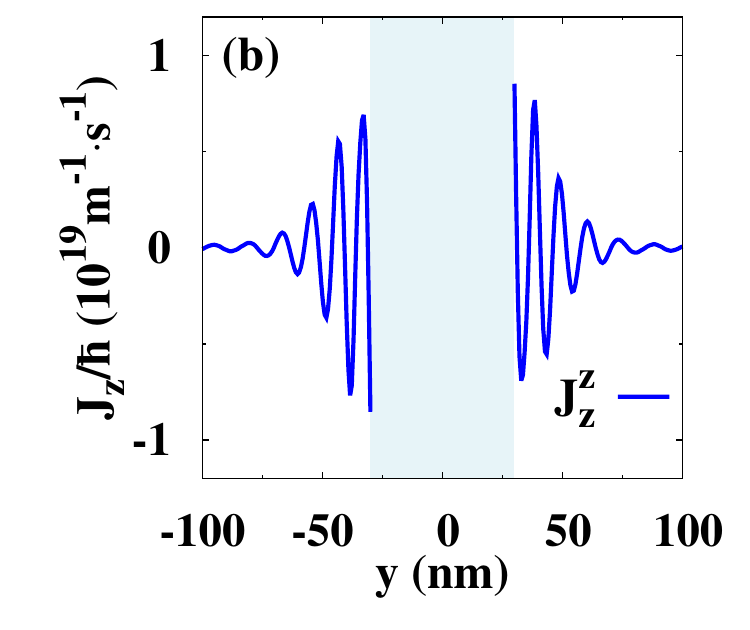}
    \end{minipage}
     \begin{minipage}[b]{0.492\columnwidth}
        \centering
        \includegraphics[width=\linewidth, clip, trim=1cm 0cm 0.2cm -1cm]{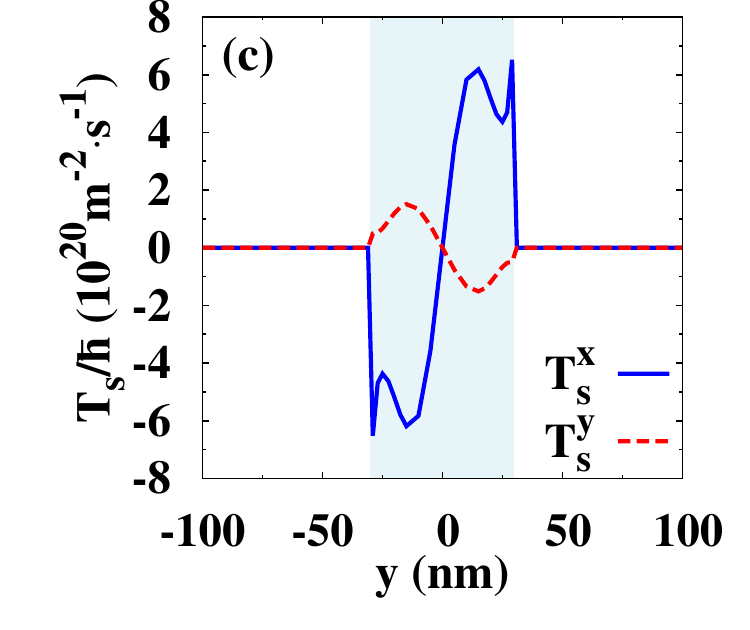}
    \end{minipage}
    \caption{Spin pumping into $p$-wave superconductor driven solely by local electric fields. The shadow region in (a)-(c) indicates the region of the near electric field.  (a) illustrates the transverse (blue arrow) and longitudinal (red arrow) spin current pumped by the one-dimensional electromagnetic field. (b) depicts the spatial profile of the transverse spin current induced by the electric field. (c) plots the spatial profile of the spin-torque density polarized along the $\hat{\bf x}$- and $\hat{\bf y}$-directions excited by the joint effect of electric and magnetic fields.} 
    \label{spin_radiation_fig}
\end{figure}

We adopt a simple implementation in which a one-dimensional ``potential barrier" is established via an optical field confined to the region $-a/2<y<a/2$ [Fig.~\ref{spin_radiation_fig}(a)], i.e.,  
\begin{align}
 {\bf E}({\pmb{\rho},t})&\simeq E_0e^{-i\omega t}\left(\Theta\left(y+\frac{a}{2}\right)-\Theta\left(y-\frac{a}{2}\right)\right)\hat{\bf z}+{\rm H.c.},\nonumber\\
 {\bf H}({\pmb{\rho},t})&\simeq {E_z}/({\mu_0c})(-i\hat{\bf x}+\hat{\bf y}),
 \label{electromagnetic_field}
 \end{align}
in which $E_0$ denotes the amplitude of the electric field and $\Theta(x)$ is the Heaviside step function, where $\Theta(x)=1$ for $x\ge 0$ and $\Theta(x)=0$ for $x<0$. This configuration describes a magnetic field rotating counterclockwise around the in-plane $\hat{\bf z}$-axis, satisfying $H_y=iH_x$, while the associated electric field oscillates along $\hat{\bf z}$.
With the electromagnetic field given in Eq.~\eqref{electromagnetic_field}, three types of nonlinear responses arise according to Table~\ref{tab:selection-rules}: (i) transverse spin current ${\bf J}^z_z$ driven by $E^*_zE_z$, and ${\bf J}^x_z$ or ${\bf J}^y_z$ driven by $E^*_zH_x$ and $E^*_zH_y$; (ii) longitudinal spin current ${\bf J}^x_y$ or ${\bf J}^y_y$ contributed by $E^*_zH_y$ and $E^*_zH_x$; (iii) spin torque ${\bf T}^x_s$ and ${\bf T}^y_s$ contributed by $E^*_zH_y$ and $E^*_zH_x$. The torque ${\bf T}_s^z$ induced by $E_z^* E_z$ vanishes in the one-dimensional geometry under consideration. Among these, the spin currents ${\bf J}^x_z$, ${\bf J}^y_z$, ${\bf J}^x_y$, and ${\bf J}^y_y$---resulting from the electric-magnetic coupling $E^*_iH_j$---are relatively small. In contrast, the transverse component ${\bf J}^z_z$,  which arises solely from the electric field ${\bf E}^2$, is expected to be the dominant contribution.
  
The temperature dependence of $\Delta_t$ is available for LaAlO$_3$/KTaO$_3$ superconducting interface~\cite{interface}, well fitted by a generalized BCS model with $T_c=2.05~\rm K$. This yields a gap magnitude $\Delta_t\approx0.36~{\rm meV}$ at $T=1~{\rm K}$. The electron density $n_e=1.8\times 10^{14}~{\rm cm}^{-2}$~\cite{interface} in the calculation corresponds to a Fermi energy $E_F\approx~431~{\rm meV}$. To ensure the non-resonance condition $\hbar\omega<2\Delta_t$, we choose the sub-terahertz photon frequency of  $\omega=0.4~{\rm THz}$~\cite{Y_Au,chirality_theorem} and the temperature $T\leq 1~{\rm K}$ with $2\Delta_t/\hbar\gtrsim 1$~THz. The electromagnetic field has an amplitude $E_0=5~{\rm kV/cm}$~\cite{optical_field1,optical_field2} and is localized in the region $a=60~{\rm nm}$.  As shown in  Fig.~\ref{spin_radiation_fig}(b), an efficient transverse spin current with polarization along the $\hat{\bf z}$-direction and flowing parallel to the electric field $\hat{\bf z}$ is induced by the electric field. The oscillation period and decay scale of ${\bf J}_z^z$ are governed by the field-region size. The proposed spin‑pumping mechanism exhibits high efficiency: the resulting spin current is comparable in magnitude to the same spin Hall current $J_z=\sigma_{zy}E_y\sim 10^{19}\hbar~(\rm{m\cdot s})^{-1}$ in systems with spin Hall conductivity of $\sigma_{zy}\sim 1.6e/(8\pi)$ under an electric field $E_y=1~{\rm kV/cm}$, which should be readily measurable according to references~\cite{efficiency1,efficiency2,efficiency3}. As shown in Fig.~\ref{spin_radiation_fig}(c), the pumped spin-torque density is polarized along the $\hat{\bf x}$-direction (governed by $E_z^*H_y$) and the $\hat{\bf y}$-direction (governed by $E_z^*H_x$) and is well localized within the pump region.

\textit{Conclusion and discussion}.---In conclusion, we have demonstrated that the triplet order parameter itself generates a wave-vector-dependent spin texture of Bogoliubov quasiparticles, giving rise to an intrinsic, non-relativistic analogue of SOC---even in the complete absence of spin–orbit interaction. Building on this foundation, we propose a mechanism to generate both spin polarization and a DC spin current with high efficiency by AC electric fields alone in triplet superconductor films. This spin transport approach may provide a definitive all-electrical spin transport fingerprint for triplet pairing symmetry, opening new pathways to rapidly and energy‑efficiently generate and manipulate spin dynamics in intrinsic triplet superconductors as well as in heterostructures exhibiting triplet superconductivity.

\begin{acknowledgments}
This work is financially supported by the National Key Research and Development Program of China under Grant No.~2023YFA1406600 and the National Natural Science Foundation of China under Grant No.~12374109. G.A.B. and I.V.B. acknowledge the support by Grant from the ministry of science and higher education of the Russian Federation No. 075-15-2025-010.
\end{acknowledgments}

\end{document}